.

# PHYSICAL ORIGINS AND LIMITATIONS OF CANONICAL QUANTUM MEASUREMENT BEHAVIOR

**Jonathan F. Schonfeld**

*Center for Astrophysics | Harvard and Smithsonian*
*60 Garden St., Cambridge MA 02138 USA*
E-mail: jschonfeld@cfa.harvard.edu

**ABSTRACT.** I review and augment my work of the last few years on the physical origins and limitations of canonical quantum measurement behavior. Central to this work is a detailed analysis of the microstructure of real measurement devices. Particular attention is paid to the Mott problem, which addresses a simpler version of canonical quantum measurement behavior: It asks why an alpha particle emitted in nuclear decay produces one and only one track in a cloud chamber. My analysis – entirely consistent with unitarity – leads to an emergent, approximate Born rule supported by experiment, with possible breakdown at very small probability density. I argue that a similar picture applies to other measurement scenarios, including Geiger counters, the Stern-Gerlach experiment and superconducting qubits.

## 1. INTRODUCTION

The purpose of this paper is to review and augment my work of the last few years (SCHONFELD, 2021, 2022, 2023, 2025) on the physical origins and limitations of canonical quantum measurement behavior. By canonical quantum measurement behavior, I mean, basically, the Born rule. I use the word "canonical" to highlight that the Born rule and related theoretical machinery are widely held to be axioms of physical law, on a level with canonical commutation relations and the gauge-field structure of electromagnetism. My results militate against this view. My work involves mathematical derivation, but it also involves mathematical conjecture. Agreement with experimental data would seem to reinforce the essential correctness of my argumentation, but I am determined to expose even the weak points so others might further advance this line of inquiry.

This is of much more than academic interest because, today, vast resources are being devoted to developing quantum computing technologies, and these appear to make heavy use of the quantum measurement axioms. As well, considerable resources are also being devoted to searching for extremely rare fundamental processes (most notably, proton decay) and it's important to know if we really can rely on canonical quantum measurement expectations when probabilities get too small.

It has been almost one hundred years since Born (BORN, 1926) proposed that the square of the absolute value of the Schroedinger wavefunction of an object being measured is proportional to the probability density of measurement. Several years later this was refined and formalized as the projection postulate (VON NEUMANN, 2018), and has been taken for granted as exact physical law ever since. Qualitatively, the gist of the postulate is that a quantum wavefunction evolves according to Schroedinger's equation (unitarity) as long as it's not being measured; but at the moment of measurement it changes discontinuously and non-unitarily ("collapses") according to application of a single, randomly manifested projection operator (with probability given by the Born rule). This puzzling dichotomy between unitary and non-unitary evolution, which has been an abiding challenge to physical intuition, is referred to as the quantum measurement problem. The effort to understand this problem in an intuitively satisfying way has given rise to a voluminous literature, very recently reviewed in (DUGIC et al, 2024) and also in (NEUMAIER, 2025). Common approaches involve interpretations of quantum ontology, or epistemology of measurement, or modifications to Schroedinger's equation. But they all have in common the assumption that Nature conforms to the projection postulate, in one form or another, to infinitely many decimal places.

To be sure, nearly a century of experimental data supports the Born rule. Every comparison between a measured scattering cross section and an ab-initio calculation tests the Born rule. Every comparison between a measured decay lifetime and Fermi's golden rule tests the Born rule. Every comparison between idealized theory and the measured pattern in a single-particle interference experiment – whether it involves photons, electrons, neutrons or large molecules (see (SCHONFELD, 2023) for references) – tests the Born rule. And yet, there is no definitive evidence that the process by which quantum mechanics selects measurement outcomes is *intrinsically* random (as opposed to reflecting the vast atomic complexity of real measurement instruments). Nor is there conclusive evidence for projection per se. Nor does the physics community maintain a systematic record of just how well the Born rule actually works quantitatively: For example, there is no tabulation of experiments by the number of decimal places in the fit to the Born rule; or by how many samples are in each measured probability bin; or by the smallest nonzero measured probabilities that are compared with the Born rule. Indeed, there are no generally agreed-upon figures of merit for quantifying the accuracy of the Born rule in any given situation, nor is there any reasoned intuition about where and how best to look for Born rule violations. There have been attempts to search for trilinear contributions to three-slit interference (JIN ET AL, 2017), or for anomalies in high-volume quantum computer calibration (BIALECKI ET AL, 2021), but these do not appear to be guided by any underlying rationale or strategy for optimizing the search.

These are important reasons why scientists must look beyond fundamental ideas about quantum measurement that take the Born rule or the projection postulate as a boundary condition. Another important reason is that both the Born rule and the projection postulate are formulated without any regard for the definition of measurement itself, much less for the internal structure of the actual measurement apparatus. This seems absurd on its face. I received my first inkling that the detailed microstructure of the apparatus cannot be disregarded when I analyzed the statistics of an over-idealized fluorescence photodetector (SCHONFELD, 2019) and found that the concept of Born's rule only makes sense at the discrete locations of the detector's individual molecules. In the present review we will go much further.

It is fair to say that all theoretical approaches that take the Born rule or the projection postulate as a given have this in common: They focus on the wavefunction of the object being measured; more succinctly, they are "object centric." They idealize, or abstract away, or – in the language of density matrices – "trace away" the microscopic details of the measurement apparatus. It seems clear to me that to really understand canonical quantum measurement behavior, and where it breaks down, one must pursue an "apparatus centric" approach that assumes neither the Born rule nor the projection postulate as axiomatic, but rather derives what an observer actually sees directly from the apparatus's microstructure. That is what I have done. In the future, comprehensive reviews of quantum foundations should cover not just object centric approaches, but apparatus centric approaches as well.

It is important to distinguish between the work reviewed here and the well-known apparatus-centric work of Allahverdyan et al (ALLAHVERDYAN ET AL, 2013). At the heart of that work is an analysis of the density matrix for a measured object (a spin). I eschew density matrices because I want to understand explicitly how a single measurement trial results in an apparently single measurement outcome.

Most of my work has focused on the cloud chamber, because it is arguably the simplest particle detector ("the 'hydrogen atom' of quantum measurement"); and because its version of the quantum measurement problem – the Mott problem – is also particularly simple.

The next section presents a theory of cloud chamber detection. Section 3 addresses experimental tests of this theory. Section 4 extends the thought process developed for cloud chambers to more sophisticated measurement scenarios involving charged particle detection. Section 5 considers possible implications for uncharged-particle measurement. Section 6 concludes with a summary and some comments on further prospects. An appendix contains a few mathematical details.

## 2. THEORY OF CLOUD CHAMBER DETECTION

A cloud chamber is an enclosure containing air supersaturated with a condensable vapor, which can be water but is more typically ethyl alcohol. When a charged particle passes through the chamber, it ionizes air molecules, and the resulting ions nucleate visible vapor droplets, which line up in a track along the particle's path. But this is only part of the story, as we shall see.

The cloud chamber's particular version of the quantum measurement problem – the "Mott problem" – refers to a very particular detection scenario: A single atomic nucleus that decays by s-wave alpha emission is placed in a cloud chamber. Everything quantum mechanical about the nuclear decay in isolation is spherically symmetric, but only one alpha detection is observed in the cloud chambers, and that's in the form of a decidedly non-spherically-symmetric track. The Mott problem asks how a spherically symmetric initial state gets manifested as a single linear track. The problem is named for N. F. Mott, who attempted to explain in quantum mechanical terms why a linear track is the generic manifestation of a charged particle in a cloud chamber (MOTT, 1929). What Mott actually showed was that the second-order perturbation-theory contribution to the wavefunction of an alpha decaying from an unstable nucleus into a gas of separated ionizable atoms is dominated by terms in which two ionized atoms line up with the source nucleus. Each such term suggests a linear track, but there are many such terms, so Mott's analysis doesn't explain why decay manifests as a single track, much less what particular feature of the detector is responsible for selecting that particular track direction. And Mott's analysis is silent about another striking feature of alpha tracks emanating from nuclear decay: Visible tracks do not originate at the nucleus itself, but rather at some nontrivial offset, on the order of centimeters (see Figure 1). What explains that, and what particular feature of the detector is responsible for selecting that particular offset?

The remainder of this section describes my findings about what really happens when an atomic nucleus decays in a cloud chamber. The picture is more complicated than previously realized. [One might suppose the picture would be similar for a bubble chamber, another track-based charged particle detector that exploits a liquid-gas phase transition. That may be correct but the energetics are quite different. See (SCHONFELD, 2021).]

The key physical ingredients that determine how, when and where a track starts – and how close it comes to any semblance of "canonical" – in the Mott scenario are as follows.

1. The alpha particle wavefunction and square-norm flux.
2. The constitution of the cloud chamber medium.
3. Singularities in ionization cross sections.

We discuss each of these in turn.

***The alpha particle wavefunction and square-norm flux***

We begin with the apparatus-free alpha wavefunction. This is important because it defines the ab initio physical interface between the object being measured (the alpha) and the measurement apparatus (the cloud chamber). Under other circumstances, it is very common for a theorist to model an initial particle state as, say, a Gaussian wavepacket; this guess-as-idealization is made entirely for the theorist's convenience. But I am very uncomfortable relying on something as arbitrary as a wavepacket concocted for my convenience when analyzing something as foundational as quantum measurement behavior. This is one reason to give research priority to the Mott problem. In the Mott scenario – a single heavy nucleus decaying to another heavy nucleus by emitting only a single alpha – one does not have to guess, one knows the apparatus-free alpha wavefunction a priori. It is a Gamow state (GARCIA-CALDERON AND PEIERLS, 1976), given by (up to a time-dependent but irrelevant phase)

$$\psi(x,t) = \theta(t - r/v)\frac{1}{r}\left(\frac{\gamma}{4\pi v}\right)^{1/2} exp\left(\left(\frac{r}{v} - t\right)\left(\frac{\gamma}{2} + i\frac{pv}{\hbar}\right)\right), \tag{1}$$

where $\theta$ is the step function, $t$ is time, $r$ is distance from the nucleus, $\gamma$ is the decay e-folding rate, $v$ is alpha speed and $p$ is alpha momentum. [This corrects a sign error in (SCHONFELD, 2021).] [Gamow states can be generalized to multi-particle decays (SCHONFELD, 2022a).] If one ignores interaction with the apparatus, then the wavefunction of the entire system is the outer product of Expression (1) and the many-degree-of-freedom wavefunction of all the cloud chamber molecules. Using the language of scattering theory (TAYLOR, 1972), it is more productive to say that this outer product is the interaction-free wavefunction of the entire system *in the un-ionized channel*. As we shall see, a detection track happens when the total square-norm of the un-ionized channel is, essentially, altogether depleted. Square-norm of an individual particle is the spatial integral of the square-norm density $|\text{wavefunction}|^2$; the total square-norm of Expression (1) – and therefore the square-norm of the interaction-free system wavefunction in the un-ionized channel – is normalized to unity, but inelastic collisions with cloud chamber molecules will tend to transfer that channel's square-norm to the ionized channel. The heart of my analysis is a detailed analysis of that square-norm transfer process.

I use the terms "square-norm density" and "square-norm" rather than the more conventional "probability density" and "probability" because I want strictly to avoid even the hint of assuming a priori that quantum mechanics is intrinsically random or probabilistic. Regardless of what one calls it, the total square-norm of a system's wavefunction is finite and conserved under unitary evolution, so the total square-norm of the un-ionized channel is a finite resource and it is entirely legitimate to examine whether and how it can be depleted.

I highlighted that Expression (1) is a Gamow state because such a state, although not stationary (up to phase), bears important formal similarity to a bound state in the discrete spectrum, and I exploit that similarity. Just as the propagator (Green's function) of a quantum particle with bound states includes a discrete term for each bound state, the propagator for a particle that can be emitted in a very slow radioactive decay includes a discrete term for the decay's Gamow state (GARCIA-CALDERON AND PEIERLS, 1976). With that in mind, I have conjectured that the following property of bound states generalizes to slow-decay Gamow states: When a bound state is eroded over time by a radiative transition (could be radiation of light or emission of an ionization electron into the continuum), the resulting wavefunction is the sum of a radiative state whose square norm increases over time, plus the original bound state wavefunction multiplied overall by a factor of magnitude < 1 that decreases with time (i.e. the bound state's square norm is not debited by eating away the bound state wavefunction piecemeal according to a spatially nontrivial pattern). This enables me to say that if the alpha wavefunction triggers ionization in the cloud chamber, then the post-ionization state gets its square-norm by debiting the entire Gamow-state wavefunction, including the part of the wavefunction that extends beyond the boundaries of the cloud chamber. I see no way around this conjecture, but it requires a more rigorous treatment than I have been able to provide. [Perhaps it relies on the limit of extremely large cross-section discussed in the last part of this Section 2.]

The post-ionization alpha wavefunction is very different from the apparatus-free Gamow state, and is another important part of the overall cloud chamber picture. When an alpha ionizes a molecule, the outgoing alpha wavefunction is basically a de-Broglie-wavelength beam radiating from an aperture of roughly molecular size [this is essentially what happens in (MOTT, 1929)]. For cases of practical interest, this beam is very narrow: For a 5 MeV alpha, the de Broglie wavelength is ~5x$10^{-15}$m, while a molecule has diameter a few x$10^{-10}$m; so the outgoing beam is a cone with opening angle < 5x$10^{-5}$ radians. For all practical purposes, this means the outgoing alpha wavefunction is perfectly collimated, because, over the typical distance – a fraction of a micron (MORI, 2014) – between subsequent droplets in a cloud chamber track, the alpha beam spreads ~$10^{-11}$m, a small fraction of its molecular-diameter width; and the next ionization will reset the beam width back to molecular diameter. Clearly, once the entire alpha wavefunction is channeled into a narrow cone emanating from a single ionization event, subsequent ionizations will keep it collimated along the same direction. So we may restate the Mott problem: What is the event that first directs (nearly) all the square-norm of the un-ionized-channel apparatus-free alpha wavefunction into a narrow cone in the ionized channel emanating from a single ionization site?

Mathematically, without identifying square norm with probability, the rate at which an interaction draws square-norm out of the un-ionized channel is determined by the cross section $\sigma$ of the interaction and the square-norm flux $\mathbf{J}$ within the channel in the vicinity of the interaction. Specifically, the rate is $\sigma\|\mathbf{J}\|$ (TAYLOR, 1972). A typical molecular cross-section is small, but, for a collimated wavefunction, square-norm is drawn from the un-ionized channel very quickly (square-norm flux is very large because the wavefunction packs a lot of square-norm into a narrow volume). In this way, ionization is perceived to take place with certainty when collimated beam and molecular ionization target are roughly within $\sigma$ of one another, and we are able to conceptualize a collimated wavefunction as a classical point particle. But square norm is not large for the apparatus-free decay wavefunction, so the interaction that collimates it must be exceptional rather than typical. We shall see shortly that the exceptions result from the dynamic constitution of the cloud chamber medium.

It will be helpful later to note here that for slow decays (small $\gamma$), high-momentum alpha particles, and times much longer than the alpha's chamber transit time, the outward square-norm flux of the apparatus-free wavefunction inside the cloud chamber at radius $r$ from the source nucleus is

$$\mathbf{J} = \frac{\hbar}{2mi}[\psi^*\nabla\psi - \psi\nabla\psi^*] \sim \mathbf{r}(p/m)|\psi|^2 \sim \mathbf{r}(\gamma/4\pi r^2)e^{-\gamma t}, \tag{2}$$

where **r** is the unit vector pointing away from the nucleus, $m$ is alpha particle mass and, obviously, $p=mv$.

### *The constitution of the cloud chamber medium*

The supercooled cloud chamber medium consists of well-separated air molecules (mostly $O_2$ and $N_2$), and vapor molecules in varying degrees of clustering. These are all targets for ionization by an alpha particle emitted by nuclear decay, and they all have similar local potential-energy environments for an electron destined to be ejected by ionization. But a cluster possesses a reservoir of energy in the form of collective molecular polarization that can offset single-molecule ionization potential. This has a major impact on the kinematics of ionization, which in turn can have a profound impact on ionization cross section, and that is what makes the difference between a "typical" ionization encounter and the exceptional one that collimates the entire apparatus-free wavefunction into a single track.

In particular, when an ion appears, say, in the center of a spherical vapor cluster, the ion's unscreened charge induces polarization in the surrounding cluster medium. The polarization response has energy

$$\frac{Q^2}{2}\left(1-\frac{1}{\varepsilon}\right)\left(\frac{1}{R}-\frac{1}{R_i}\right), \tag{3}$$

where $Q$ is ion charge, $\varepsilon$ is cluster dielectric constant, $R$ is cluster radius and $R_i$ is an effective radius of the volume that the ion itself occupies in the cluster. The dielectric constant may be close to unity for open air, but is much greater than one for a vapor cluster. Therefore, Expression (3), manifestly a negative number, is easily large enough to compete with the binding energy of the electron that had to get ejected to produce the ion in the first place. For appropriate values of $R$, or for an analogous criterion in a differently shaped cluster, Expression (3) can even offset the binding energy exactly. In that case an alpha particle can ionize a vapor molecule with no energy loss. As we discuss below, this is a singular case in quantum Coulombic scattering, and leads to anomalously large ionization cross sections that can indeed collimate apparatus-free wavefunctions

Of course we must be quantitative about what makes an ionization cross section "anomalously large enough." For this reason, in (SCHONFELD, 2021) I introduced a parameter $\tau$ to represent the evaporation time of a vapor cluster that initiates a track. Ionization cross section $\sigma$ for a molecule in a cluster is "large enough" when square-norm flux **J** flowing through $\sigma$ drains nearly all the square-norm from the apparatus-free wavefunction before the cluster in question can evaporate. Mathematically, for unit overall normalization, that amounts to $\sigma\|\mathbf{J}\|\tau > 1$.

[In (SCHONFELD, 2021) I stated the following puzzle: If there's a collection of nuclei all close to one another in a cloud chamber, why don't they all produce alpha decay tracks at the same time when "the right" subcritical cluster comes along? I.e. why don't we see multiple tracks issuing from the same starting point at the same time? There are several factors working against this. First, even if the nuclei in the collection are close to one another, they're not at exactly at the same location, and therefore $\|\mathbf{J}\|$ won't be the same for all of them at the cluster in question. And second, once a first electron is emitted to form the first ion in a given cluster, the next emitted electron – if there is going to be one – would experience a different binding environment because of the ionic charge exposed by the first emitted electron.]

It is convenient to talk about clusters as if they are randomly distributed in a continuous sample space with probability $\rho$ per unit radius $R$, per volume of location and per unit time of occurrence. But continuity cannot be strictly correct, because a cluster is built out of finitely many molecules. Indeed, I have estimated (SCHONFELD, 2025) that a cluster at the starting points of a cloud chamber track has ~25 molecules. This will become important below when we talk about limits to the Born rule. Probabilistic randomness also cannot be strictly correct, but it must be a pretty good idealization (with or without unitarity) because there are so many molecules of all kinds in a macroscopic cloud chamber. Certainly the reader must not make the mistake of thinking that I am somehow assuming *quantum* randomness a priori.

### *Singularities in ionization cross sections*

A tendency toward singularity is built into quantum Coulombic cross sections because of the long-range nature of the Coulomb potential (TAYLOR, 1972). Certainly the cross section singularity of elastic Coulomb scattering is well known. Singularity also is generic in inelastic Coulomb scattering if there is another state that is degenerate with the initial state of the target molecule (TAYLOR, 1972). That is not the case in typical laboratory experiments, but it happens at a vapor cluster when polarization energy (Expression (3)) exactly cancels the binding energy of an electron that would be expelled from a molecule by ionization and that would be emitted at zero kinetic energy. Thus, as the radius $R$ of a cluster approaches a value $R_c$ for which cancellation is exact, total ionization cross section approaches the singular form $A/(R_c\text{-}R)$ for some coefficient $A$. [This echoes another Coulombic singularity discussed much earlier in (FEINBERG et al, 1986), although never verified experimentally, to my knowledge.]

From this expression, and the distribution $\rho$ of cluster radii, we immediately obtain a statistical distribution of track starting points in space and time (SCHONFELD, 2021). Specifically, we arrive at the expression $\rho_c A v \tau |\psi(\mathbf{x},t)|^2$ for the spatial-temporal probability density (per unit volume and unit time) for track initiation at detector location $\mathbf{x}$ and time $t$, where $\psi$ is the apparatus-free wavefunction and $\rho_c$ is the value of $\rho$ at $R=R_c$. This is a Born rule, in that it's a probability density proportional to the absolute-value squared of a wavefunction. This particular rule for cloud chambers is completely novel. For it to really be a Born rule in the full sense of the quantum measurement axioms, one would have to reframe cloud chamber detection as measurement of alpha particle position, and to accept that the definitive signature of alpha position is the starting point of its track. Such a reorientation of perspective is entirely unanticipated in the literature. It's not clear to me if or how it relates to the projection postulate, because it's not at all clear whether projection operators exist that could select for track starting points. In any event, this Born rule can be turned into a particular functional form by applying Equation (1), for times much longer than the alpha's chamber transit time:

$$\rho_c A v \tau |\psi(\mathbf{x},t)|^2 = \frac{1}{r^2}\frac{\rho_c A \gamma \tau}{4\pi} e^{-\gamma t}. \tag{4}$$

This is what we will compare with experiment in Section 3. [See (SCHONFELD, 2021) for a discussion of how this is consistent with Bell's theorem.]

It is very important to highlight why this Born rule can only be approximate: First, because it refers to a wavefunction at a particular spatial position $\mathbf{x}$ even though the anomalous cross section is undoubtedly large enough to encompass a considerable range of other positions. Second, because not all clusters are spherical and not every vapor molecule that is a target for ionization is located at the center of a cluster. Third, because the expression for ionization cross section near critical radius assumes a continuum idealization for the cluster medium, whereas a real cluster is a kind of "stick figure" consisting of a relatively small number of vapor molecules. Fourth, because it neglects possible effects of many small-cross-section elastic interactions with air and vapor molecules that can accumulate over the long time that elapses for a slow decay, and modify the wavefunction in the no-ionization channel.

It is especially important to highlight that this Born rule is obtained without explicit reference to any non-unitary process. Indeed, the argumentation is built around the continuous flow of conserved square-norm, which is a necessary (but not sufficient) signature of unitarity. Nevertheless, I am sometimes asked whether there might be other ways in which my argumentation is circular, i.e. whether I somehow implicitly assume quantum measurement axioms in trying to explain physically how apparently axiomatic behavior comes about. Clearly, I assume that condensed matter exists (otherwise there could be no cloud camber), and that, at a microscopic (or at least mesoscopic) level, it exhibits statistical behavior characterized by stable distributions. However, I do not derive these assumptions themselves from the most basic unitary first principles. So I suppose there may be some opening for circularity. Still, it is a highly nontrivial result to deduce a previously unknown, very specific distribution (4) from other assumptions that are based on what we know actually does happen, and that do not seem even to foreshadow, let alone resemble anything like the Born rule. Perhaps it is too early to say that I have *derived* a Born rule, but I think I can fairly claim to have *explained* one.

## 3. EXPERIMENTAL TEST OF CLOUD CHAMBER THEORY

No one has ever conducted a cloud chamber experiment with the express purpose of measuring the distribution of starting points of tracks from decaying nuclei. In view of the Born rule prediction in Equation (4), this is a glaring gap in the literature and needs addressing. Perhaps this gap is to be expected, since there had previously been little awareness that a track *doesn't* start right at the decaying nucleus itself.

However, since the advent of the Internet, educators have posted many pedagogical videos of cloud chambers in operation, and some of these videos contain images that are relevant to testing Equation (4). For reasons explained in (SCHONFELD, 2022), the most suitable such video [22] was produced by Jefferson Lab in 2010.

This video shows 90 seconds of alpha decay tracks materializing in a cloud chamber containing a needle source of $^{210}$Pb. A needle source is literally a sewing needle whose tip has been thinly plated with a radioactive species. $^{210}$Pb is itself a beta emitter, but decays to $^{210}$Po, an alpha emitter with half-life of several months. The chamber is a Petri dish, supercooled by sitting on a bed of dry ice. The camera is placed directly above the flat lid of the dish. I examined every video frame and measured the coordinates at which every track started. From that data I compiled the cumulative distribution function (CDF) of track start positions as a function of distance from the source in the plane of the image. I also modeled what the same CDF would turn out to be if the Born rule (4) governed the statistics of track start locations in three-dimensional space. This also involved data fitting in order to fix the unknown parameter $\rho_c A \tau$ in Equation (4). The details of this modeling and fitting can be found in (SCHONFELD, 2022).

Figure 1 shows a representative frame from the Jefferson Lab video, in which the reader can see that tracks originate away from the tip of the needle. Figure 2 shows the measured CDF and the best fit model based on the Born rule prediction.

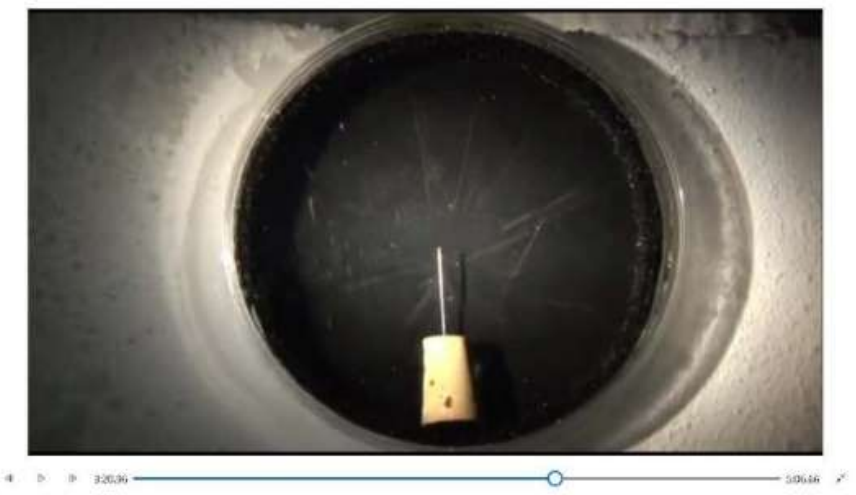

Figure 1. Frame #6030 from [22].

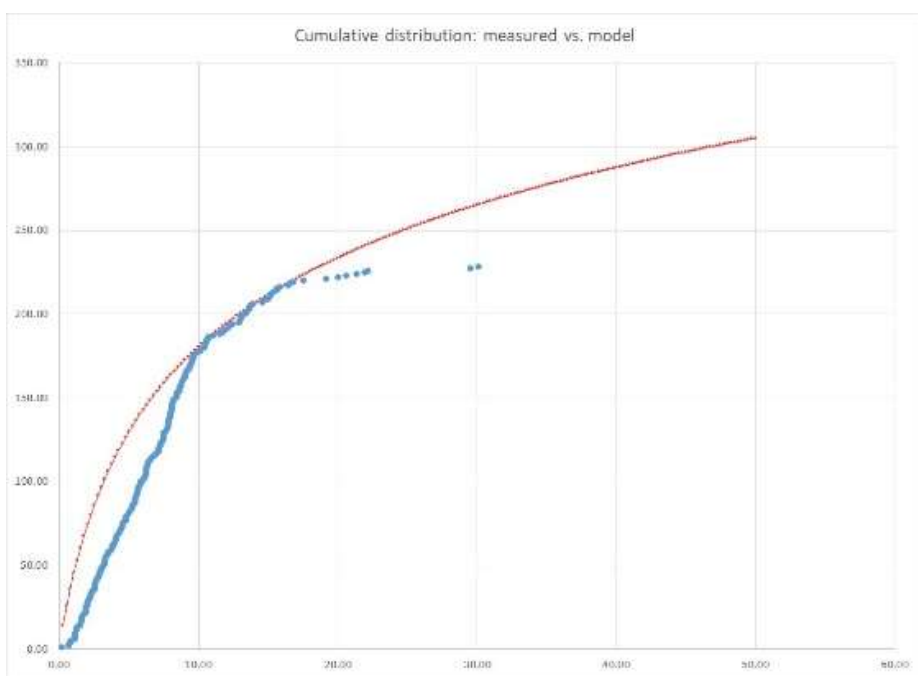

Figure 2. Measured (blue) and Born-rule modeled (red) CDF. Vertical axis calibrated in total counts, horizontal axis calibrated in mm.

For radii up to ~20mm, the quality of the fit ranges from mediocre to good. For radii beyond ~20mm there is a marked deficit of detections in the measured data compared with the model. In (SCHONFELD, 2022) I explored several candidate explanations for this apparent deficit, primarily based on considering the distribution of heat in and around the cloud chamber. I was able to rule out some, and found none of the rest compelling. [There also seems to be a shortfall in detections at radius below roughly 10mm; this may or may not be an artifact of the fitting procedure. I considered and rejected the possibility that it came about because of local heating by the radioactive source itself.] I speculated that the deficit beyond ~20mm might indicate a breakdown of the Born rule at small wavefunction. Detection for very small wavefunction might require a vapor cluster with |R-Rc| much smaller than could actually be achieved with a stick-figure made of finitely many vapor molecules. I have not been able to exclude this possibility.

The quality of the fit in Figure 2 is a test of the functional form of the Born rule (4), and the apparent lack of detections beyond ~20mm is a possible signature of the underlying microphysics. The normalization of the fit would also provide a test if one could draw on a robust a priori theory of the coefficient $\rho_c A\,\tau$ for comparison. There is a literature aimed at quantitatively predicting the statistical distribution of cluster sizes in a supercooled vapor, but it is not developed enough to enable a really robust comparison with our experimental data. In (SCHONFELD, 2025) I adapted results from (BAUER ET AL, 2001) to validate the measured value of $\rho_c A\,\tau$ to within a factor of 1-2. This is impressive when one considers how the source tables in (BAUER ET AL, 2001) range over many orders of magnitude.

The analysis of opportunistic video data in this section provides some encouragement that the cloud chamber theory in Section 2 is on the right track. Clearly, there is a strong need for much more careful and more fine-grained experimentation. We also require a much more robust quantitative understanding of the transient cluster population in supercooled vapors.

## 4. GENERALIZING FROM CLOUD CHAMBERS

Ideally, we should immediately apply lessons from the cloud chamber to other, more standard measurement scenarios. However, I proceeded more cautiously, first considering an alternative charged-particle track detector in order to learn how to adapt the cloud chamber thought process to different media. In particular, it occurred to me that a Geiger counter detecting a nuclear decay might be productive as an incremental step up in sophistication, because it has strong unappreciated commonality with cloud chambers. First, both rely on alpha particles making ionization tracks. In a cloud chamber, the tracks are directly visible when we see nucleated vapor droplets; in a Geiger counter, the tracks are indirectly visible when we see voltage pulses at the internal anode. Second, both involve media in which ionization cross sections are small. In a cloud chamber, that's the air between vapor clusters; in a Geiger counter, that's the air outside the Geiger-Muller tube, and also the buffered noble gas inside. So both cloud chamber and Geiger counter have a Mott-like problem: determining precisely where the apparatus-free alpha wavefunction gets collimated. As we have seen, the answer for a cloud chamber is that collimation takes place at exceptional vapor clusters that come and go due to thermal fluctuation. It seems intuitively obvious that, for a Geiger counter, collimation has no choice but to take place at or inside the very thin mica window where particles enter the metallic tube, because none of the gases inside or outside the Geiger-Muller tube support clustering of any kind, let alone exceptional clusters with very large ionization cross sections (and the metal surface of the tube is too far away and the thin-wire anode is too narrow). I conducted an experiment to test this hypothesis (SCHONFELD, 2023).

There are actually three hypotheses here: Collimation takes place (i) at the interface between window and air, (ii) at the interface between window and buffered noble gas, and (iii) in the interior of the window. Since writing (SCHONFELD, 2023), I have learned that my experiment can provide critical discrimination between the three cases. This is important because the three cases have very different microphysical implications. In case (iii), collimation must take place at conjectured localized interior structures (voids or crystal defects) where collective polarization produces energetics very similar to Equation (3) (although now the molecule to be ionized must be at the edge of a void rather than somewhere in the interior of a cluster). In cases (i) and (ii), Equation (1) is replaced by basically the same thing (up to overall scale) without the $1/R$ term (like the electrostatic potential of a point charge over a conducting plane). In these cases, the value of $R_i$ is still related to the radius of the ionization target, but is also sensitive to the topographical scale of surface roughness.

The three hypotheses also differ significantly in another way that one can exploit for experimental analysis. In case (i), an alpha particle collimated at the air-window interface experiences slowing all the way through the window due to interaction with mica before it comes out the other side, in the interior of the tube, where ions can create charge avalanches at the anode. In case (ii), an alpha particle collimated at the buffered-ideal-gas-window interface experiences no slowing due to mica. In case (iii), the amount of slowing depends on precisely where in the mica the collimation takes place. I have modeled the experiment under these three pictures, and find that only case (i) appears to track the data satisfactorily. [In (SCHONFELD, 2023), I hadn't thought through this diversity of cases, and unwittingly modeled only case (i) (and I did so with an unrelated error that turned out not to have a significant impact on the results).]

Of course all these cases aren't mutually exclusive and in reality could operate in tandem. But the agreement of case (i) with the data does seems not to cry out for blending with the other cases. This makes some intuitive sense, because as the alpha wavefunction propagates from the decaying nucleus, it encounters the air-window interface first, and then the interior of the window, and finally the buffered-ideal-gas-window interface. I.e. we can at least satisfy ourselves that a winner-take-all case (i) is consistent with causality, because it comes first in the propagation sequence

In my experiment (see Figure 3), I mounted a needle source of the alpha emitter $^{210}$Po on a manual linear stage. I dialed the stage until the "hot" end of the needle just touched the center of the mica window of a commercial Geiger-Muller tube. I then dialed the stage back through equally spaced stops, and at each stop I recorded the Geiger counter count rate. (See (SCHONFELD, 2023) for experimental details, including data conditioning.)

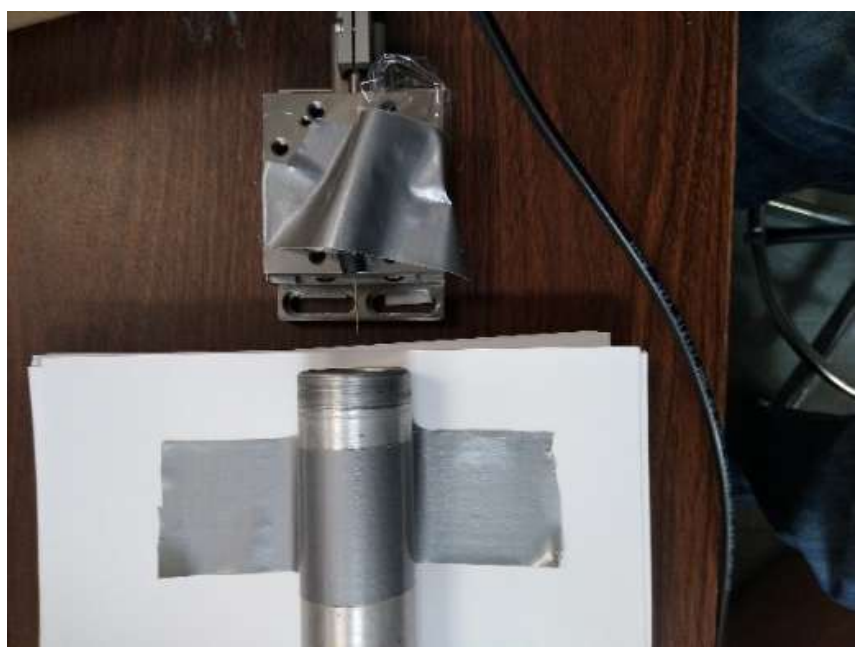

Figure 3. Geiger counter experiment. The Geiger-Muller tube is the cylinder in the bottom half of the photograph. One can make out the very thin source needle issuing from the plug at the lower edge of the movable stage in the top half of the photograph. From (SCHONFELD, 2023).

Figures 4-6 compare the experimental data with models based on cases (i)-(iii), as well as with a naive "geometric" model that assumes collimation takes place at the decaying nucleus itself. [This latter model is basically geometric flux through the window, moderated by attenuation in air and mica.] The central fact of each model is that alphas only make it into the Geiger-Muller tube if the total slowing distance (in air and window for geometric model, and in window only for cases (i)-(iii)) is equivalent to less-than-complete stopping in air. The equations for these models are described in the Appendix. In the model for case (i), I set the value of window thickness (in equivalent slowing-distance in air) to optimize fit to the data. The value that I got, 16mm, is reassuringly close to the manufacturer's estimate. I assumed the same value for the other cases, but I performed some numerical variation to satisfy myself that I couldn't improve fit meaningfully. All curves are normalized to give the same value of counts per minute for zero separation between source and window.

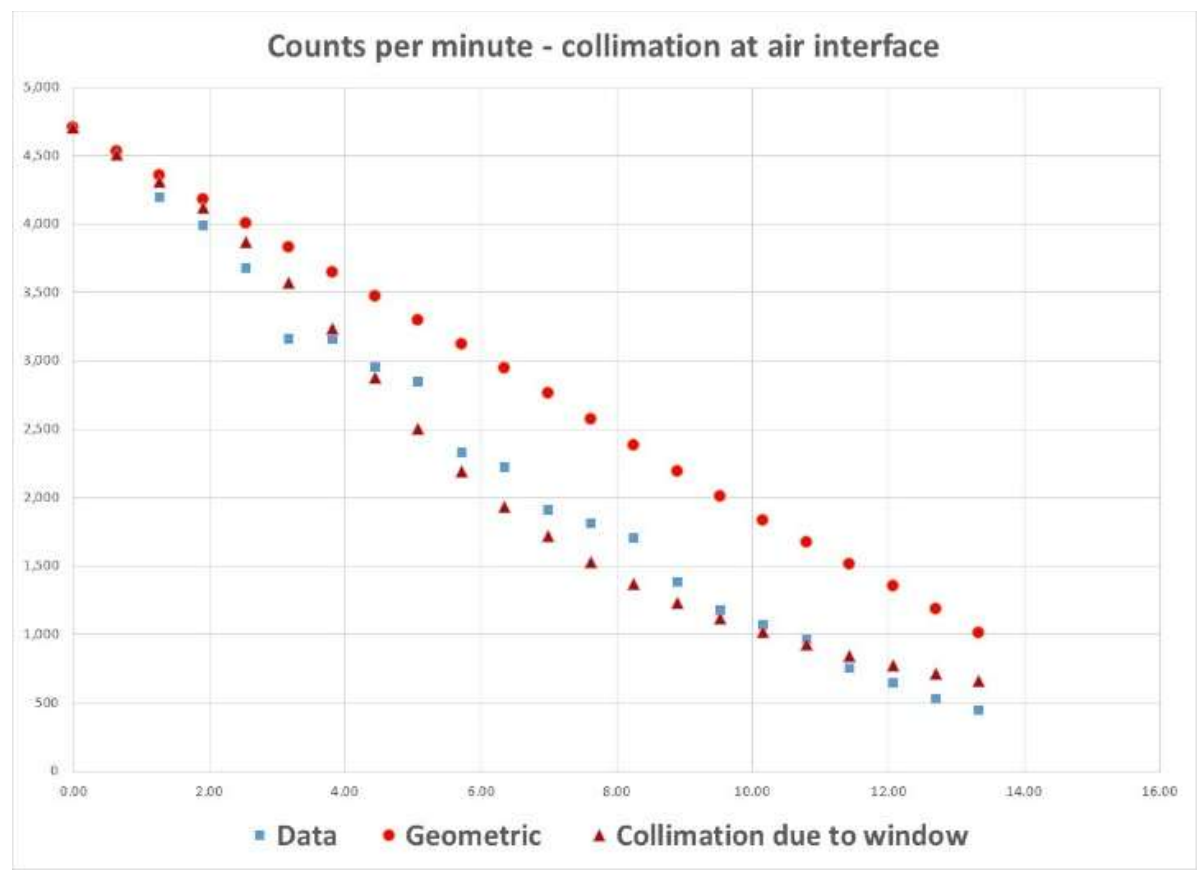

Figure 4. Data and models for Geiger counter experiment. Red circles correspond to geometrical model, purple triangles to case (i) model, and blue squares to data. Radioactive source is modeled as extending over a few millimeters.

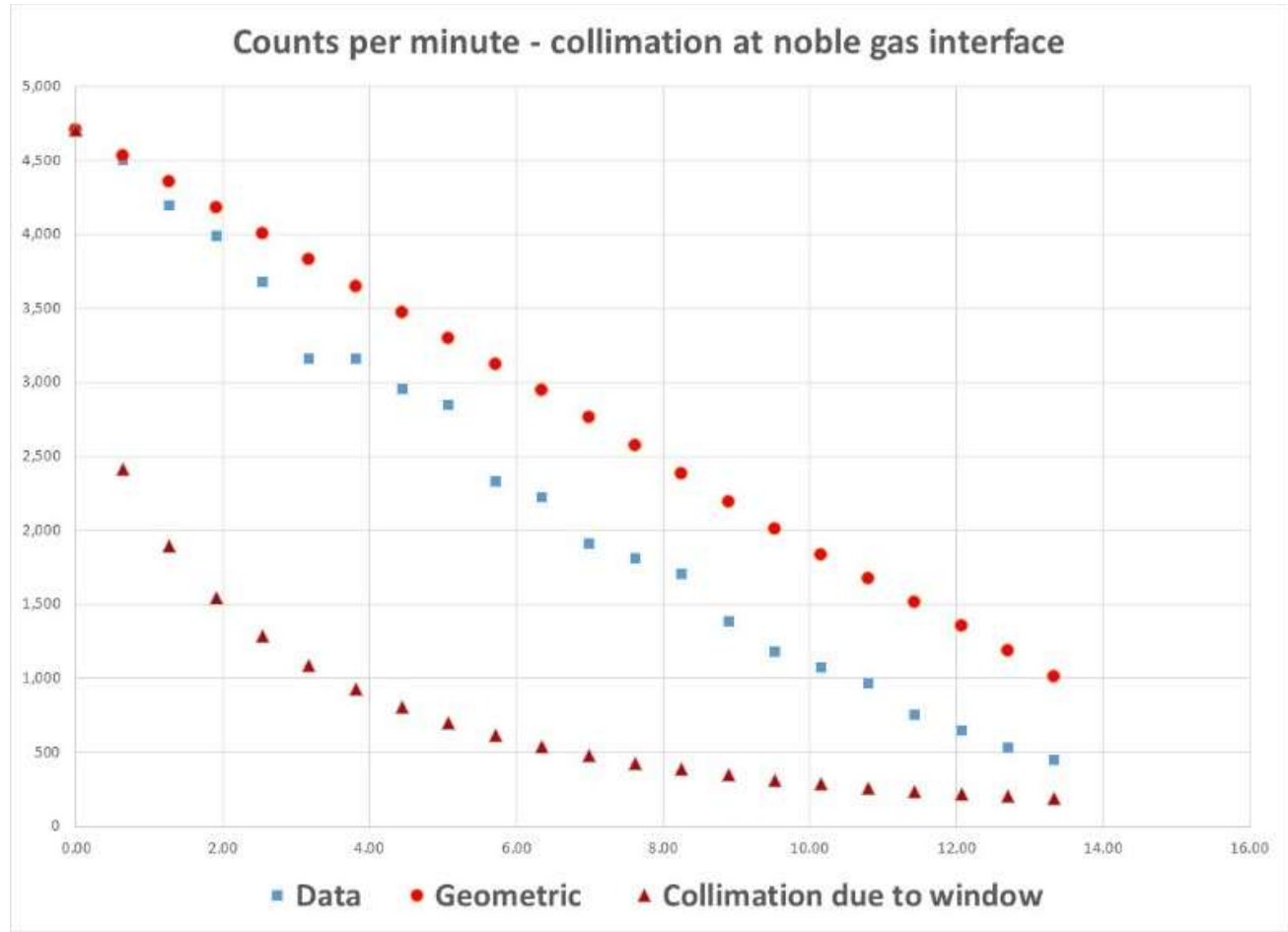

Figure 5. Data and models for Geiger counter experiment. Red circles correspond to geometrical model, purple triangles to case (ii) model, and blue squares to data. Radioactive source is modeled as extending over a few millimeters.

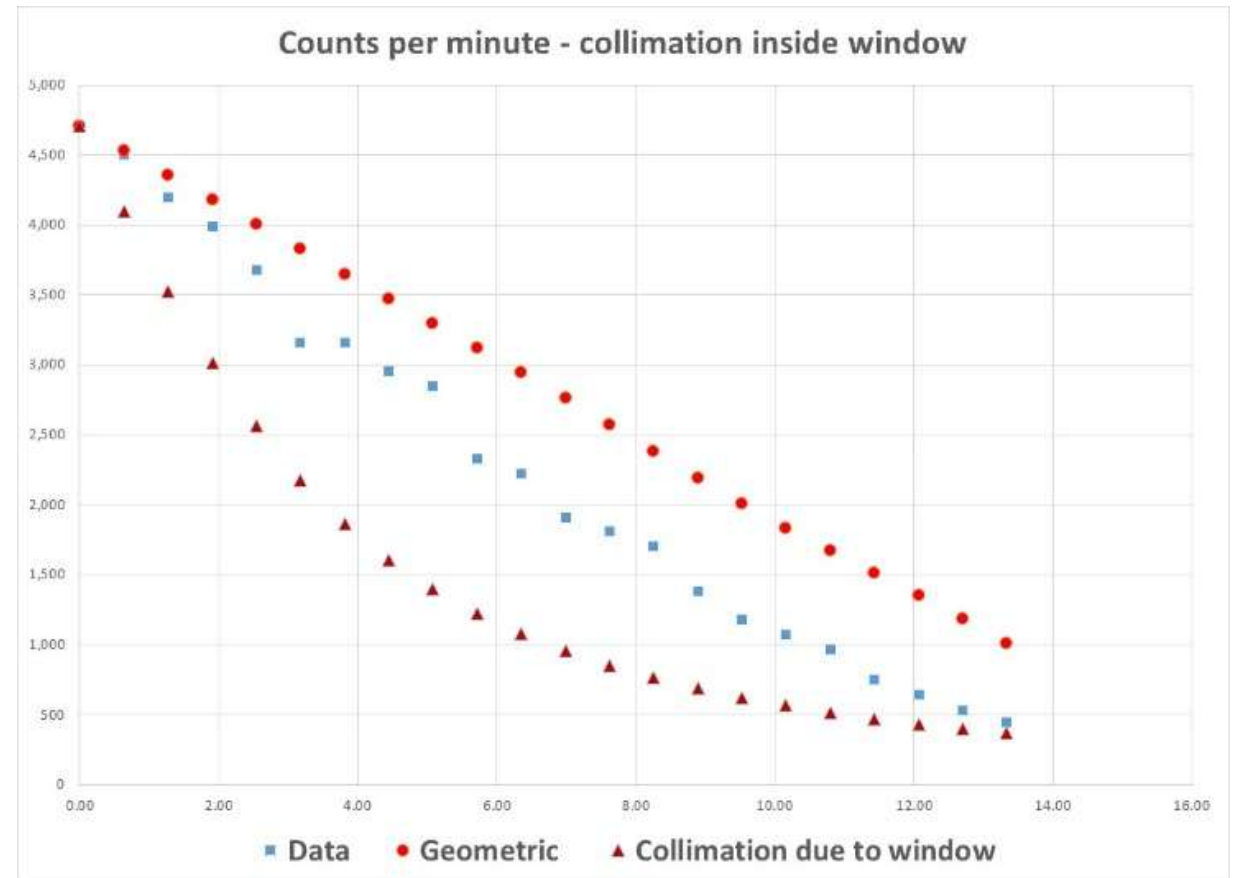

Figure 6. Data and models for Geiger counter experiment. Red circles correspond to geometrical model, purple triangles to case (iii) model, and blue squares to data. Radioactive source is modeled as extending over a few millimeters.

So, from this experiment – which needs to be repeated in a much more controlled setting – we learn that a solid state surface can collimate a single-particle wavefunction. With this recognition, it is a relatively modest step from Mott-like scenarios to more familiar quantum measurement situations. In the remainder of this section, we summarize how this applies to two specific such scenarios, the Stern-Gerlach experiment and measurement of a superconducting qubit.

In the original Stern-Gerlach experiment (GERLACH, STERN, 1922), a beam of neutral spin-1/2 silver atoms passed through a spatially varying magnetic field. The beam was then detected as two very small and close but distinct smears deposited on a glass plate. Here we assume the spin-1/2 beam particles are charged, because we are generalizing from the theory of charged particle detection in a cloud chamber set out in Section 2. Actually, this scenario is a bit hypothetical, because the Stern-Gerlach experiment is in fact much more difficult with charged than neutral particles, and as late as 2019 had not been carried out definitively (HENKEL ET AL, 2019). But it is a good stepping stone to the second scenario – superconducting qubit measurement – which is an object of active engineering development and is not at all hypothetical.

For a spin-1/2 charged particle in such a beam, the magnetic field as usual splits its wavefunction into two very small and close but distinct branches corresponding to the two cardinal spin states (up/down) of which the particle is a superposition. Both branches propagate towards an ionizing detector with a solid acceptance window. If, as my Geiger counter results indicate, the entrance interface of this window contains atoms or molecules with anomalously large ionization cross sections, then a detection probability density analogous to $\rho_c A v \tau |\psi(\mathbf{x},t)|^2$ applies here as well, where $\mathbf{x}$ is now position in the plane of the detector. And since the two wavefunction branches should be shaped identically, but separately scaled by the complex weights of the spin superposition, this immediately turns into the Born rule for spin measurement (with all the caveats about approximate validity that we saw in the cloud chamber case).

[As discussed in (SCHONFELD, 2023), this adaptation from cloud chamber theory involves some nuance. In the cloud chamber case, the Gamow state of the decay provided a single repository of square-norm available to flow in its entirety into the ionized channel. In the Stern-Gerlach case the concept of Gamow state doesn't necessarily apply. Instead, we have to assume that the anomalous ionization cross-section, wherever it occurs, is large enough to encompass the combined transverse spreads of the two closely-spaced wavefunction branches. With this in mind, it might be interesting to do a Stern-Gerlach experiment with the two branches allowed to drift very far apart before detection. Perhaps that would produce a violation of the Born rule.]

As discussed in (SCHONFELD, 2023), we can understand the Born rule in superconducting qubit measurement via the same mechanism, because there is an implicit Stern-Gerlach apparatus embedded deeply in the qubit readout system. In detail, a superconducting qubit is an artificial atom made from Josephson junctions coupled to a radio frequency (RF) resonating cavity, typically configured so that the lowest two excited states are close in energy and can be treated together as a self-contained two-level system. In dispersive readout, the state of this two-level system is measured by sending a microwave pure tone at the cavity via a transmission line, and recording the reflected signal (BLAIS ET AL, 2004). If the frequency of the pure tone is chosen appropriately, the signal reflected from one qubit basis state (up or down) has a phase shift that is detectably different from the phase shift due to reflection from the orthogonal basis state. The reflected signal goes through several stages of amplification and then passes through an analog-to-digital converter (A/D), after which it is recorded as a digitized voltage time series. The phase shift is extracted from the time series via traditional I/Q processing, and the measured state is inferred directly from the result.

All along this signal chain, except for the very last step, there is simply no opportunity for a large-cross-section event to rapidly divert square-norm into an ionization channel or something similar. It's different at the A/D converter. An A/D is basically a succession of elementary solid-state devices that determine whether an input voltage is above or below a pre-set threshold. Each such device is basically a layered semiconductor. One layer is a source of electrons, and the other layers pull these electrons one way or another depending on whether voltage is above or below the threshold. When an electron is drawn into one of the layers, it promotes another electron in the same layer into the conduction band (a sort of ionization process), and that in turn produces a kind of charge cascade observed as a voltage pulse, much as happens in a Geiger counter. In other words, the A/D is basically a set of embedded Stern-Gerlach apparatuses, and so if the Born rule applies to Stern-Gerlach because of wavefunction collimation at the detector interface, then it should apply equally to superconducting qubit measurement because of wavefunction collimation at the interfaces between semiconductor layers (again with all the caveats about approximate validity that we saw in the cloud chamber case). In a recent paper (SCHONFELD, 2025a) I extrapolated from cloud chamber experimental results to estimate the scale of expected Born rule violation in superconducting qubit measurement. I concluded that this violation might be very difficult to observe.

## 5. DETECTION OF UNCHARGED PARTICLES

So far we have focused on detection of charged particles, but of course famous signatures of canonical quantum measurement behavior are also observed with detectors of uncharged particles. These include the glass plate used to detect the neutral silver atoms in the original Stern-Gerlach experiment; photomultiplier tubes, CCDs and photographic film used to detect photons in optical slit experiments; and $BF_3$ detectors used to detect neutrons in neutron slit experiments. I am unable to comment on fogging of glass plates. In (SCHONFELD, 2023) I speculated that the boundary collimation that seems to take place at a Geiger counter window also operates in CCDs and photographic grains. In (SCHONFELD, 2021) I speculated that wavefunctions of liberated electrons are collimated at conducting surfaces in photomultiplier tubes, where the energetics resembles Equation (3). Presumably something similar applies to the charged fragments that result from the fission of a boron nucleus after it absorbs an incoming neutron in a $BF_3$ detector.

## 6. SUMMARY AND FURTHER PROSPECTS

In this paper I have reviewed work of the last few years – both theoretical and experimental – that supports the following picture of what physicists have come to think of as canonical quantum measurement behavior.

- In a cloud chamber, a visible track (and therefore a detection) appears from a charged particle with a diffuse wavefunction (such as a nuclear decay product) where and when there is an exceptional condensed vapor cluster with extremely large ionization cross section. The cross section is large because the energy of induced polarization in the cluster nearly compensates for the binding energy of the ejected electron. The statistics of track origination follows a Born rule, at least approximately.
- In a Geiger counter, an internal track (and therefore a detection) appears from a charged particle with a diffuse wavefunction (such as a nuclear decay product) where and when there is an exceptional feature with extremely large ionization cross section on the near surface of the mica entrance window. The cross section is large because the energy of induced polarization in the mica nearly compensates for the binding energy of the ejected electron. Analysis of experimental data on count rate as a function of distance from a nuclear source indicates that the statistics of track origination follows a Born rule, at least approximately.

- In a charged particle Stern-Gerlach experiment, spin measurement – with the Born rule (at least approximately) – happens when there is an exceptional feature with extremely large ionization cross section on the near surface of the entrance window to an ionizing detector. The cross section is large because the energy of induced polarization in the window nearly compensates for the binding energy of the ejected electron.
- In a superconducting qubit system, state measurement – with the Born rule (at least approximately) – happens when there is an exceptional feature with extremely large ionization cross section at a boundary between semiconductor layers in an electronic A/D converter. The cross section is large because the energy of induced polarization in the semiconductor nearly compensates for the binding energy of the ejected electron.

The common themes are clear: large ionization cross sections at exceptional features in apparatus media, related to balance between induced polarization and electron binding energy, and emergence of an approximate Born rule. I have suggested that the same themes are at play in scenarios involving measurement or detection with uncharged particles.

In (SCHONFELD, 2023), I suggested that these themes could possibly help us understand a bigger question: Why do almost all particles we observe in everyday life seem to follow tracks as if they were classical? Perhaps, in the primordial past, the wavefunctions of particles we deal with every day first became collimated by interacting with exceptional interface features of emerging condensed matter.

## APPENDIX: WAVEFUNCTION COLLIMATION AT GEIGER COUNTER WINDOW

In this appendix, I explain the equations that underlie the model curves in Figures 4-6. To do this I extend calculations that first appeared in Section 3 of (SCHONFELD, 2023). To start, refer to Figure 7.

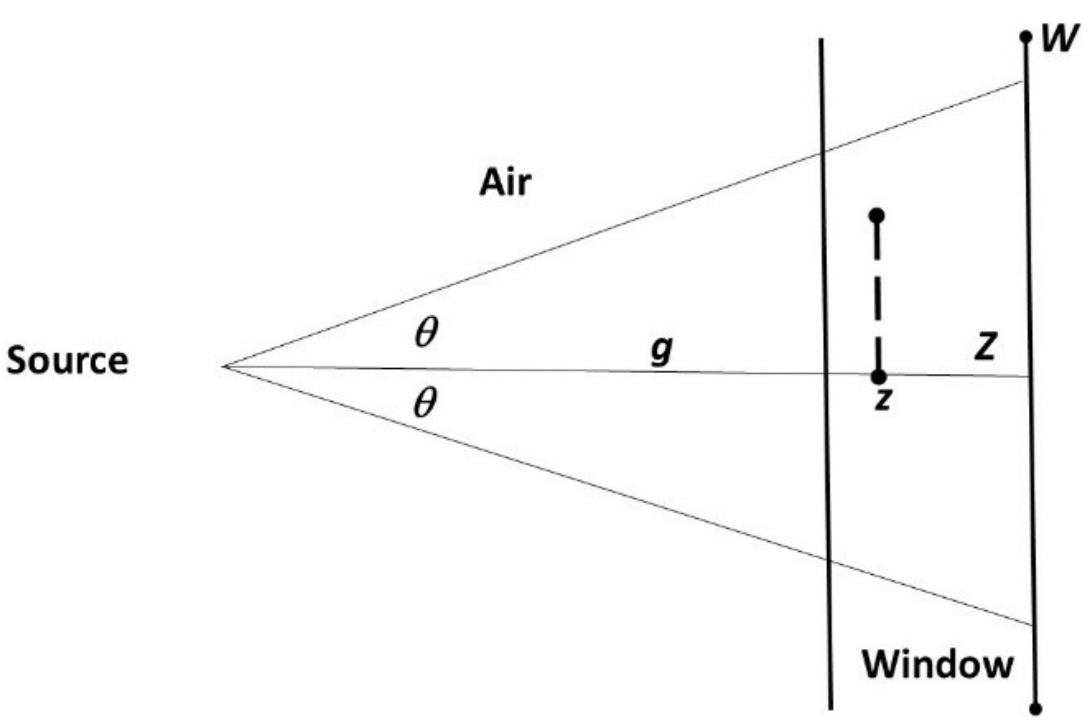

Figure 7. Geiger counter reference diagram for alpha stopping calculations.

In this diagram, $g$ is distance from radioactive source to the window's near surface, $Z$ is window thickness, $z$ is horizontal coordinate of an arbitrary point in the window measured from the near surface ($y$, length of dashed line, not labeled, is its radial coordinate [vertical in the plane of the diagram), , $W$ is window radius, and $\theta$ is the smaller of the angle of a ray which just touches the radius of the window's far surface, or (for geometric and case (i)-(ii) models) along which an alpha is just barely stopped due to a combination of slowing in air and in the window medium. For a very thin window, $z$ and $Z$ are negligible when compared to $g$ or $W$. Let $L$ be the stopping distance of an alpha particle in air, and let $S$ be a scale factor so that one unit of distance inside the window results in the same slowing as $S$ units of distance in air.

For a geometric model, in which collimation takes place at the source, alpha flux through the window is proportional to the solid angle subtended by a cone of opening angle $\theta$, i.e. proportional to $(1-\cos\theta)$. And $\theta$ itself is the smaller of $\arctan(W/g)$ (neglecting $Z$ relative to $g$) and $\arccos((g+SZ)/L)$.

For the case (i) collimation model, $\theta$ is the smaller of $\arctan(W/g)$ and $\arccos(SZ/L)$. [I assume alpha slowing doesn't happen until the wavefunction is collimated (I assumed the opposite in (Schonfeld, 2023)). Otherwise there would be no tracks in video [22] that terminate more than ~40mm from the source, regardless of where they start. But in fact some such tracks can be seen.] For the case (ii) model, $\theta$ is simply $\arctan(W/g)$ because there is no slowing until the alpha has fully entered the Geiger-Muller tube. Following Equation (4), the alpha flux into the detector for cases (i) and (ii) is proportional to the integral of $1/((g+z)^2+y^2)$ over the relevant window surface, i.e. proportional to $\ln(\cos\theta)$.

For case (iii), alpha flux into the detector is the integral of $1/((g+z)^2+y^2)$ over the window interior, but only including any point $(z,y)$ for which the ray that traces back to the source then exits the window and satisfies the constraint that distance to the far window surface along the same ray has length less than $L/S$. Calculating alpha flux into the detector in this model is elementary and only the end result is given here: For $g>W(SZ/L)/(1-(SZ/L)^2)^{1/2}$, flux is proportional to $(SZ/L)\ln((g^2+W^2)/g^2)$. For $g<W(SZ/L)/(1-(SZ/L)^2)^{1/2}$, flux is proportional (with the same multiplier) to

$$\left[\frac{SZ}{L}-\frac{g}{(g^2+W^2)^{1/2}}\right]-\frac{SZ}{L}\ln\frac{SZ}{L}. \tag{5}$$

**Acknowledgments**

I am grateful to Miroljub Dugic (University of Kragujevac) for suggesting this submission. I am grateful to Jay Lawrence (Dartmouth University, University of Chicago), Arnold Neumaier (University of Vienna) and Li Hua Yu (Brookhaven National Laboratory) for valuable dialogue.